\documentclass[useAMS,usenatbib]{mn2e}
\usepackage{times}
\usepackage{graphicx}
\newcommand{\ltsimeq}{\raisebox{-0.6ex}{$\,\stackrel
           {\raisebox{-.2ex}{$\textstyle <$}}{\sim}\,$}} 

\title[Near-infrared evolution of Sakurai's Object]{Sakurai's Object: characterising the near-infrared CO ejecta between 2003 and 2007}

\author[H. L. Worters et al.]{H. L. Worters$^{1,2}$,
M. T. Rushton$^{1}$,
S. P. S. Eyres$^{1}$,
T. R. Geballe$^{3}$
A. Evans$^{4}$\\ 
$^1$Centre for Astrophysics, University of Central Lancashire,
Preston, PR1 2HE, UK\\
$^2$South African Astronomical Observatory, Observatory, 7935, South Africa\\
$^3$Gemini Observatory, 670 N. A'ohoku Place, Hilo, HI 96720, USA\\
$^4$Astrophysics Group, Keele University, Keele, Staffordshire ST5 5BG, UK}
\date{Received; in original form}

\begin{document}
\label{firstpage}
\maketitle
\begin{abstract}
We present observations of Sakurai's Object obtained at $1-5 \mu$m
between 2003 and 2007. By fitting a radiative transfer model to an
echelle spectrum of CO fundamental absorption features around $4.7
\mu$m, we determine the excitation conditions in the line-forming
region.  We find $^{12} \rm C/^{13} \rm C=3.5^{+2.0}_{-1.5}$,
consistent with CO originating in ejecta processed by the very late
thermal pulse, rather than in the pre-existing planetary nebula.  We
demonstrate the existence of $2.2 \times 10^{-6} \leq M_{\rm CO} \leq
2.7 \times 10^{-6}$~M$_{\odot}$ of CO ejecta outside the dust, forming
a high-velocity wind of $500 \pm 80$~km~s$^{-1}$.  We find evidence
for significant weakening of the CO band and cooling of the dust
around the central star between 2003 and 2005.  The gas and dust
temperatures are implausibly high for stellar radiation to be the sole
contributor.
\end{abstract}


\section{Introduction}
\label{sec-intro}
Sakurai's Object is a highly evolved post-AGB star that had begun to
venture down the white dwarf cooling track when, in 1995, it underwent
sudden rebrightening \citep{Nak96} due to a final helium shell flash,
or very late thermal pulse (VLTP) \citep{Due97}.  Since then,
Sakurai's Object has undergone observable changes on timescales of
weeks to months, providing an instance in which this very brief stage
of evolution experienced by $\sim 15\%$ of intermediate-mass stars
\citep{Ibe96} could be tracked over a period of only a few years.
Several phases of dust production followed the outburst, with a deep
optical minimum beginning in early 1999, such that any changes in the
central star have since been inferred from radio and infrared
observations \citep{Tyn00, Haj05, Van07}.  Subsequent
observations and modelling have revealed much about the dust shell
formation and the outer regions of the ejecta
[e.g. \citet{Tyn02,Kim02}].

Using observations at 2.3~$\mu$m from 1998, \citet{Pav04} modelled
overtone CO band absorption in the stellar atmosphere to determine a
$^{12}$C/$^{13}$C ratio of $4\pm 1$, consistent with VLTP
nucleosynthesis.  In order to follow the development of the ejecta,
annual monitoring of the target in the near-infrared has been
undertaken.  The discovery of fundamental band (4.7~$\mu$m) lines of
CO in the wind was reported by \citet{Eyr04}, based on low resolution
spectroscopy obtained in 2002 and 2003.  Here we present the analysis
of observations of the CO fundamental lines in data obtained between
2003 and 2007, mostly at higher resolution.

\section{Observations}
\label{sec-observations}
All data presented here were obtained at the United Kingdom Infrared
Telescope on Mauna Kea, Hawaii, using the facility spectrographs CGS4
and UIST. A summary of the observations is given in Table 1. Low
resolution spectra were predominantly obtained with UIST, utilising
the $0.24^{\prime\prime}$ and $0.48^{\prime\prime}$ slits, but in one
instance (2005 August 3) CGS4 and its $0.61^{\prime\prime}$ slit were
used.  For all observations the telescope was nodded in the standard
ABBA pattern. Standard stars were observed just prior to or after
Sakurai's Object and at airmasses close to those of the target. The
spectra of Sakurai's Object were ratioed by those of the appropriate
calibration stars (Table 1).  Flux calibration was performed utilising
photometry of the standard stars and colour corrections based on
\citet{Tok00}.  Wavelength calibration was obtained from an arc lamp
or from telluric absorption lines in the spectrum of the calibration
star, and was accurate to better than 0.0001~$\mu$m for the UIST and
low resolution CGS4 spectra.
 
In the case of the CGS4 echelle observations on 2004 June~10, three
echelle settings were used with a slit width of $0.9^{\prime\prime}$
to cover the wavelength range 4.68-4.77~$\mu$m at a resolving power of
20000 (15~km~s$^{-1}$). Following data reduction, the three spectra
were adjoined after applying small scaling factors to two of the
spectra in order to match the continuum levels.  Wavelength
calibration for these spectra was derived from telluric CO lines and
is accurate to 3~km~s$^{-1}$.

\begin{table*}
\begin{tabular}{lcclll}
\centering
Date &  $t_{\rm int}$ &  Waveband & $R$ & \multicolumn{2}{c|}{Calibration Star}\\
     & (s)           &  ($\mu$m)   &     & Name & Spectral Type\\
\hline
2003-09-08  & 720 &   1.40-2.51  & 500  & HIP~86814 & F6V\\
2003-09-08  & 480 &   2.91-3.64  & 1400 & BS~6378 & A2IV-V\\
2003-09-08  & 240 &   3.62-4.23  & 1400 & BS~6378 & A2IV-V\\
2003-09-08  & 960 &   4.38-5.31  & 1200 & BS~6378 & A2IV-V\\
2004-06-10  & 1200 &  1.40-2.51  & 2000 & HIP~86814 & F6V\\
2004-06-10  & 840  &  2.23-2.99  & 2000 & HIP~86814 & F6V\\
2004-06-10  & 1440 &  2.91-3.64  & 2000 & BS~6378 & A2IV-V\\
2004-06-10  & 720  &  3.62-4.23  & 2000 & BS~6378 & A2IV-V\\
2004-06-10  & 1120 &  4.38-5.31  & 2000 & BS~6496 & F5V\\
2004-06-10$^{\ast}$ & 600 & 4.678-4.692  & 20000 & BS~6378 & A2IV-V \\
2004-06-10$^{\ast}$ & 600  & 4.713-4.727 & 20000 & BS~6378 & A2IV-V\\
2004-06-10$^{\ast}$ & 510 & 4.758-4.762  & 20000 & BS~6378 & A2IV-V \\
2005-08-03$^{\ast}$ & 560 & 4.48-5.12 & 2000 & BS~6595 & F5V \\	
2007-06-02  & 1080 &  3.62-4.23 & 2000 & BS~6595 & F5V \\	
2007-06-02  & 2400 &  4.38-5.31 & 2000 & BS~6595 & F5V \\	
\end{tabular}
\label{tab-obs}
\caption{Log of near-infrared spectroscopic observations obtained with
  UKIRT and presented in this paper.  An asterisk denotes an
  observation made with CGS4; all remaining spectra were acquired
  using UIST.}
\end{table*}

\section{Results}
\label{sec-results}

\begin{figure}
    \includegraphics[width=0.99\columnwidth]{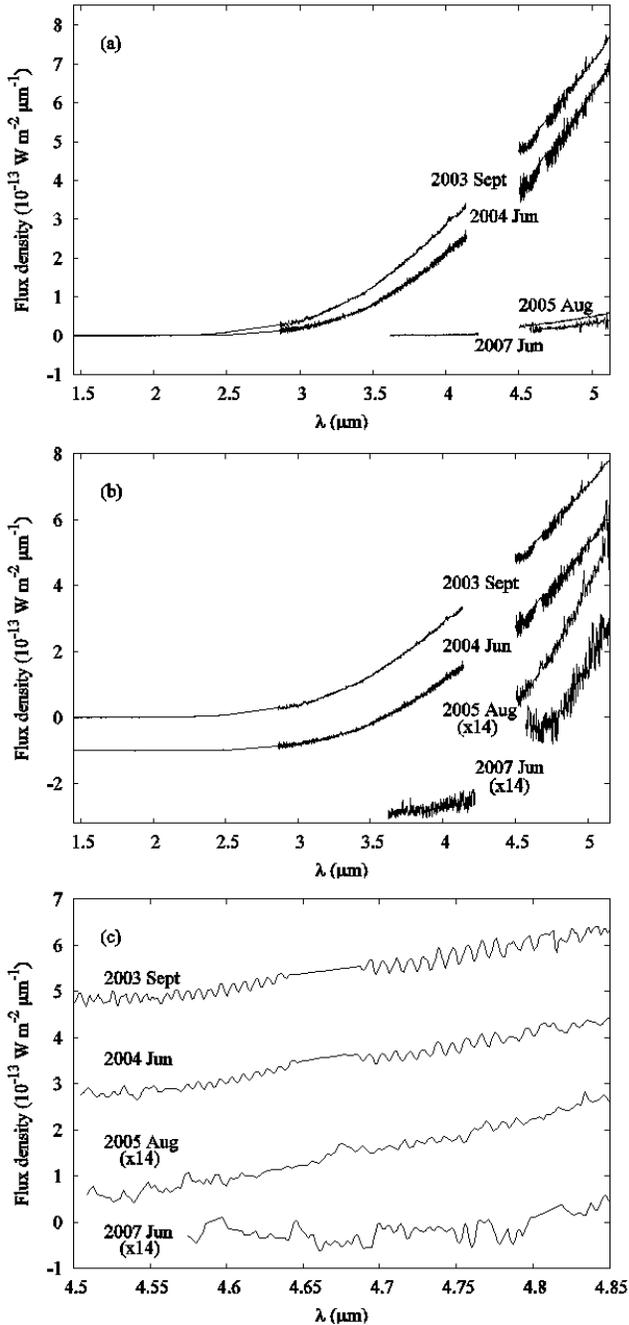}
    \caption{(a) UKIRT spectra of Sakurai's Object on 2003 September
      8, 2004 June 10, 2005 August 3 and 2007 June 2.  Gaps occur in
      the data due to strong telluric absorption.  (b) The same UKIRT
      spectra shown in Fig.~\ref{fig-co1-5}\@(a), with 2005 and 2007
      spectra scaled up by a factor of 14 to aid comparison of
      continuum shapes.  Spectra from 2004, 2005 and 2007 have been
      offset vertically by $-1$, $-3$ and $-3 \times
      10^{-13}$~W~m$^{-2} \mu$m$^{-1}$, respectively. (c) An
      enlargement of Fig.~\ref{fig-co1-5}\@(b) over the spectral range
      of the $^{12}$CO P-branch and $^{13}$CO R-branch lines.  The
      2003 and 2004 spectra are interpolated between 4.64 and $4.68
      \mu$m because of two strong hydrogen absorption lines in the
      spectrum of the A2 calibration star.}
    \label{fig-co1-5}
  \end{figure}

Figure~\ref{fig-co1-5}\@(a) shows spectra of Sakurai's Object in the
$1-5 \mu$m region from 2003 to 2007.  The continuum can be seen to
fade from one year to the next, with a substantial decrease in flux
density between the 2004 and 2005 observations.  To aid comparison of
the continuum shapes, Figure~\ref{fig-co1-5}\@(b) shows the 2005 and
2007 spectra scaled up by a factor of 14.  Vertical offsets have also
been applied to the 2004, 2005 and 2007 spectra for display purposes.
A difference in gradient of the continua can be seen, particularly in
the M-band, indicating a blackbody peak at increasing wavelength in
consecutive years, strongly suggesting cooling of the dust.  The only
features superimposed upon the continuum are the fundamental
absorption lines of CO around $4.7 \mu$m.  These can be seen in
Figure~\ref{fig-co1-5}\@(c); an enlargement of the $^{12}$CO P-branch
and $^{13}$CO R-branch region.  The 2004, 2005 and 2007 low resolution
spectra have been convolved with a Gaussian profile of full width at
half maximum (FWHM) equal to the resolution of the 2003
spectrum. Comparing the spectra, one can see a significant weakening
of the CO band from 2003 to 2005.  Although individual lines cannot
clearly be seen in the 2005 spectrum, the apparent emission bump at
$4.67 \mu$m, the wavelength of the CO band centre, is a strong
indication that weak P- and R-branch absorption lines are still
present.  The low signal-to-noise ratio of the 2007 spectrum precludes
a meaningful estimate of the CO band strength.

\begin{figure}
    \includegraphics[width=0.99\columnwidth]{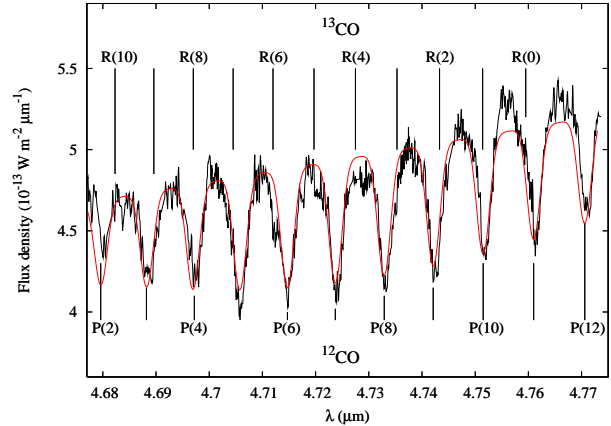}
    \caption{UKIRT echelle spectrum of CO (1-0) absorption in
      Sakurai's Object from 2004 June 10.  $^{12}$CO P-branch and
      $^{13}$CO R-branch lines are labelled; the wavelengths of the
      lines are for a stellar radial velocity of $+115$~km~s$^{-1}$
      and peak absorption at a heliocentric velocity of
      $-270$~km~s$^{-1}$.  The smooth red line superimposed upon the
      spectrum corresponds to a spherical shell model spectrum
      consisting solely of $^{12}$C$^{16}$O.}
   \label{fig-echelle}
  \end{figure}

Figure~\ref{fig-echelle} is the echelle spectrum from 2004 June,
showing $^{12} \rm CO$ and $^{13} \rm CO$ in absorption around $4.7
\mu$m.  The absorption features have non-Gaussian profiles and show
remarkably large linewidths.  We measure a full width at zero
intensity (FWZI) of $400 \pm 70 $~km~s$^{-1}$ and FWHM of $180 \pm
15$~km~s$^{-1}$.  By measuring the blueshift of absorption features in
the echelle spectrum we can improve upon the $290 \pm 30$~km~s$^{-1}$
absorption maximum found by \citet{Eyr04} using observations made at
lower resolution.  After correcting for the heliocentric radial
velocity of +115~km~s$^{-1}$ \citep{Due96}, we find the velocity of
the absorption maximum to be $270 \pm 15$~km~s$^{-1}$ relative to the
star.

\section{Modelling the CO}
\label{sec-modelling}

In order to ascertain whether the CO detected outside the dust in 2004
originates in ejecta from the central star observed by \citet{Pav04},
or is associated with the old planetary nebula (PN) or interstellar
medium, we estimate the $^{12} \rm C/^{13} \rm C$ isotopic ratio by
applying a simple radiative transfer model to the $4.68-4.78 \mu$m
echelle spectrum, shown in Figure~\ref{fig-echelle}.  Because the
absorption features are extremely broad, the $^{13}$CO lines are
blended with the stronger $^{12}$CO lines. Their presence has a
tangible effect on the spectrum.  There is an apparent deepening of
the $^{12}$CO absorption features where $^{13}$CO lines are coincident
[e.g. $^{12}$CO P(5) and $^{13}$CO R(7)], while the perceived
continuum level is depressed in regions where $^{13}$CO lines occur in
between $^{12}$CO lines [e.g. around $^{13}$CO R(4)].  To illustrate
this point, the smooth red line in Figure~\ref{fig-echelle} depicts a
model spectrum for the case of absorption by $^{12}$CO alone,
superimposed upon the echelle spectrum.

Primarily, we model the CO assuming a spherical geometry.  For
comparison, we consider an isothermal, plane parallel slab of
absorbing CO as a secondary model.  The results presented here are for
the spherical model, favoured because it considers emission as well as
absorption integrated over all lines of sight from Sakurai's Object,
rather than a single pencil beam of light.

\subsection{Spherically symmetric shell model}
\label{sec-sphere}

The CO absorption region is modelled as an expanding, spherically
symmetric shell, centred on the star. Assuming dust obscuration
prevents the gas from seeing the central star, we take the radiation
source to be the intervening dust.  The validity of this assumption is
discussed in Section~\ref{sec-geometry}. By fitting a blackbody
function to the $1-5 \mu$m continuum, we find dust temperatures of
$360 \pm 30$~K in 2003, and $320 \pm 40$~K in 2004, consistent with
the value of $350 \pm 30$~K estimated by \citet{Eyr04}. The radiative
transfer model is insensitive to temperature because of the limited
wavelength coverage of the echelle spectrum.  We therefore fix the
temperature of the dust at $320$~K.  The relative strengths of the low
and high $J$ $^{12}$CO lines place constraints on the gas temperature
of $320 \pm 30$~K.  We fix this parameter in the model accordingly.

The strength of the absorption component is determined by solving the
radiative transfer equation along lines of sight within the shell.  We
employ the CO fundamental linelist taken from \citet{Goo94}, including
vibrational levels up to $v^{\prime\prime}=8$.  All identified lines
correspond to $v=1-0$ lines, indicating no significant population of
higher vibrational levels.  The average measured FWHM of the CO lines
is 180~km~s$^{-1}$. We use this value to constrain velocity broadening
in the model.

The model is optimised for column density $N({\rm CO})$ and $^{12} \rm
C/^{13} \rm C$ isotopic ratio, and is fitted to the data using a
minimum $\chi^{2}$ test.  The output value corresponding to the lowest
$\chi^{2}$ was established by use of the downhill simplex method of
\citet{Pre92}.  Uncertainties in the output were estimated from the
range of values produced by several iterations of the model, varying
the starting points of the free parameter values.  The output
calculated by the model for the free parameters is given in Table~2,
and compared with results from previously published works on Sakurai's
Object.  An isotopic ratio of $^{12} \rm C/^{13} \rm C =
3.5^{+2.0}_{-1.5}$ and column density $N({\rm CO}) =
(7.6^{+1.5}_{-0.5}) \times 10^{16}$~cm$^{-2}$ are determined by the
fitting procedure.  These output values were obtained after 48
iterations of the model.  The $\chi^{2}$ value that was achieved, 5.0,
implies a relatively poor fit.  Fixing the continuum temperature
resulted in some difficulty in modelling the data, likely due to
non-local thermal equilibrium population of the energy levels.  A
blackbody curve is a reasonable approximation to the dust continuum;
slight deviation may be due to instrumental effects.

There is a factor of 10 difference between the column density of CO in
2004, as determined by the spherical shell model, and that estimated
for 2003 by \citet{Eyr04} (Table~2).  This is partially due to the
weakening of the CO band between 2003 and 2004
[Figure~\ref{fig-co1-5}\@(c)].  The 2003 estimate assumed an
isothermal slab of absorbing CO, and is consistent with the value
$N({\rm CO}) \approx 6 \times 10^{17}$~cm$^{-2}$ determined by our
plane parallel slab model for 2004.  In part, we attribute the lower
column density to the fact that the spherical model integrates the
absorption over all lines of sight from Sakurai's Object, hence a
larger value is expected when a single line of sight is used to
characterise the column density, as in the isothermal slab model. The
two models find consistent $^{12}\rm C/^{13}\rm C$ isotopic ratios,
independent of the geometry.

\begin{table*}
\begin{tabular}{lccc}
Parameter & Model values & Published values & Date of published observations \\
\hline	       		    
Dust temperature (K) & 320 (fixed) & $350 \pm 30 ^{\ast}$ & 2003 \\
Velocity broadening (km~s$^{-1}$) & 180 (fixed) & $25-200^{\ast}$ & 2003\\
CO excitation temperature (K) & $320 \pm 30$ (fixed) & $400 \pm 100 ^{\ast}$ & 2003\\
Column density ($\times 10^{16}$ cm$^{-2}$ ) & $7.6^{+1.5}_{-0.5}$ & $70^{+30 \ast}_{-20}$ & 2003\\
$^{12}$C/$^{13}$C & $3.5^{+2.0}_{-1.5}$ & $1.5-5 ^{\dag}$, $4 \pm 1 ^{\ddag}$, & 1996, 1998 \\ 
& &$\geq 3 ^{\ast}$, $3.2^{+3.2 \S}_{-1.6}$ & 2003, 2005 \\
\end{tabular}
\label{tab-par}
\caption{Values for each parameter of the spherically symmetric shell
  model, applied to the 2004 echelle spectra.  Column two shows values
  calculated by the model, compared with values from the literature in
  column three.  Because of the variable nature of Sakurai's Object,
  column four gives the date of the observation to which each value
  from the literature corresponds.  The dust temperature is determined
  by fitting a blackbody function to the continuum and is fixed in the
  model, while the temperature of the gas is determined using relative
  strengths of CO lines.  The velocity broadening is constrained by
  the average FWHM of CO lines in the echelle spectra.  $^{\ast}$Eyres
  et al. (2004). $^{\dag}$Asplund et al. (1999). $^{\ddag}$Pavlenko et
  al. (2004). $^{\S}$Evans et al. (2006).}
\end{table*}

\section{Discussion}
\label{sec-discussion}

\subsection{$^{12} \rm C/^{13} \rm C$ ratio}
\label{sec-12c13c}

Possible contributors to $^{13}$C observed in the $^{13}$CO
fundamental band lines are the interstellar medium, the pre-existing
PN and the post-VLTP ejecta from Sakurai's Object.  The low $^{12} \rm
C/^{13} \rm C$ ratio and high velocity of the material modelled here
preclude origins in the ISM [$^{12} \rm C/^{13} \rm C \sim 70$
\citep{She07}] or the evolved PN [$20 \leq ^{12} \rm C/^{13} \rm C
\leq 40$ \citep{Pal00}, $v \sim 15$~km~s$^{-1}$ \citep{Ack93}].
\citet{Pav04} fitted synthetic spectra to UKIRT echelle spectra of
Sakurai's Object from 1998, modelling the cool stellar atmosphere in
the $2.32-2.38 \mu$m region.  Using this technique they find an
isotopic ratio of $^{12}$C/$^{13}$C $=4\pm1$ in material close to the
star, consistent with the VLTP interpretation of Sakurai's Object.
Here we find a compatible $^{12}$C/$^{13}$C ratio in material observed
outside the dust shell in 2004.

\begin{figure}
    \includegraphics[width=0.99\columnwidth]{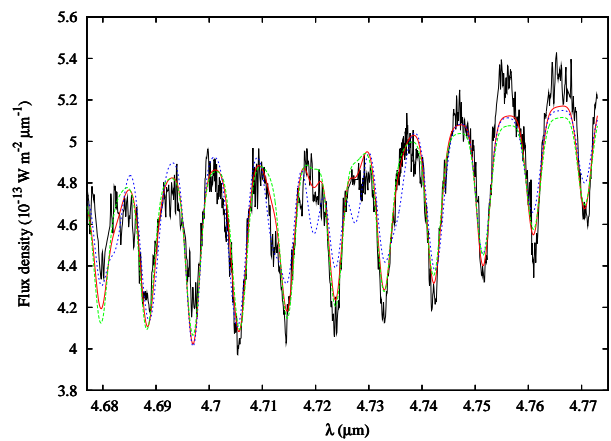}
\caption{Spherically symmetric shell model CO spectra superimposed upon
the 2004 June UKIRT echelle spectrum.  Plots correspond to
$^{12}$C/$^{13}$C ratios of 10 (green dashed line), 1 (blue dotted
line) and the best fit of $^{12}$C/$^{13}$C = 3.5 (solid red line).}
\label{fig-ratios}
\end{figure}

Figure~\ref{fig-ratios} shows model spectra produced by the
spherically symmetric shell model described in
Section~\ref{sec-sphere} for different $^{12}$C/$^{13}$C ratios,
superimposed upon the echelle spectrum.  The best fit found by the
model is depicted by the solid red line, for a $^{12}$C/$^{13}$C ratio
of 3.5.  The blue dotted line corresponds to $^{12}\rm C/^{13}\rm C =
1$, in which the $^{13}$CO lines are more pronounced than observed in
Sakurai's Object (R(4) and R(5) in particular), whereas in the green
dashed line, corresponding to $^{12}\rm C/^{13}\rm C = 10$, the
absorption features are weaker than observed and there is
insufficient depression of the continuum level at wavelengths where a
$^{13}$CO line is expected.  The isothermal slab model is in agreement
with the best-fit result of the spherically symmetric model, finding
$^{12} \rm C/^{13} \rm C \approx 3$.

\subsection{CO velocity}
\label{sec-geometry}
From the echelle spectrum we measure a FWHM of the CO absorption
features corresponding to a velocity broadening of $180 \pm 15
$~km~s$^{-1}$.  A gas shell expansion velocity of $500 \pm
80$~km~s$^{-1}$ is obtained from the difference between the mean
radial velocity of the blueward edge of the CO lines ($-385 \pm
80$~km~s$^{-1}$) and the systemic radial velocity
($+115$~km~s$^{-1}$).

The absorption maximum occurs at $-270 \pm 15$~km~s$^{-1}$, while the
mean FWZI measures $400 \pm 70$~km~s$^{-1}$.  Thus, assuming
spherically symmetric expansion, material contributing to the redward
edge of the line is moving at $-70 \pm 50$~km~s$^{-1}$ with respect to
the observer, i.e. with a velocity of $500 \pm 80$~km~s$^{-1}$ away
from the central star, at an angle close to the plane of the sky.
Similarly, the blueward edge ($-470 \pm 50$~km~s$^{-1}$) corresponds
to gaseous motion along the line of sight to the observer.  The fact
that we see no emission from the limbs or the far side of the shell
supports the model configuration of a star surrounded by concentric
shells of dust and gas; the inferior dust shell obscures the far side
of the superior gas shell, hence we detect only the blueshifted gas
along the line of sight and do not see the redshifted (P Cygni)
emission.

\subsection{Evolution of the CO}
\label{sec-co}

In 1998 March, broad, blueshifted absorption of He~I ($1.083 \mu$m),
which was absent from a spectrum of equal resolution obtained in 1997
July, was detected around Sakurai's Object \citep{Eyr99}.  We assume
the He~I and CO-bearing materials were mobilised contemporaneously.
Taking an expansion velocity of 500~km~s$^{-1}$ from the time of He~I
detection places limits on the outer gas shell radius of $9.9 \times
10^{10} \leq R_{\rm CO} \leq 1.1 \times 10^{11}$~km in 2004 June;
increasing to $1.5 \times 10^{11} \leq R_{\rm CO} \leq 1.6 \times
10^{11}$~km at the time of the most recent observation in 2007.  This
recent gas radius is comparable with the range on the dust shell
radius of $4.3 \times 10^{10} - 1.5 \times 10^{11}$~km, modelled by
\citet{Van07}, and approximately 17~times larger than the 1997 dust
radius of $8.4 \times 10^{9}$~km determined by \citet{Kip99}.  The
consistent temperatures of CO and the continuum source (Table~2)
indicate at least partial mixing of the gas and dust; we therefore
take the radius of the CO shell to be a good approximation for that of
the dust shell.

Shock acceleration to $500$~km~s$^{-1}$ would result in dissociation
of the CO.  The absence of CO in the spectrum of Sakurai's Object in
1999 could be used to support a hypothesis of CO destruction in the
1998 onset of the fast wind, recombining by the 2000 April observation
\citep{Eyr04}.  Alternatively, we suggest gentle acceleration of the
gas by the dust due to radiation pressure.  In this scenario it is
expected that some of the dust would precede and hence obscure the
gas; the CO only becoming visible with expansion and thinning of the
dust shell over time.

Assuming a thin, spherically symmetric gas shell, and taking a column
density of $7.6 \times 10^{16} $~cm$^{-2}$ (derived by the spherically
symmetric shell model), we place limits on the total CO ejecta mass of
$4.4 \times 10^{24} \leq M_{\rm CO} \leq 5.4 \times 10^{24}$~kg
(i.e. $2.2 \times 10^{-6} \leq M_{\rm CO} \leq 2.7 \times
10^{-6}$~M$_{\odot}$), dependent on the date the wind commenced.
Significant asymmetry or asphericity of the nebular geometry
[e.g. \citet{Van07,Kim08}] would cause this value to vary.

\subsection{Dust temperature}
\label{sec-dust}

The equilibrium temperature of the dust at the radius determined in
Section~\ref{sec-co} is $\sim 100$~K.  This is based on assumptions of
1~$\mu$m graphitic carbon grains \citep{Tyn02}; a stellar luminosity
of 3000~L$_\odot$ \citep{Her01,Tyn02} and an effective temperature of
5200~K \citep{Pav02}.  An equilibrium temperature of 320~K at this
radius would require an impossibly high luminosity; a factor of $>100$
greater than predicted by the models of \citet{Her01} and
\citet{Tyn02}.  The CO temperature of 320~K is also inconsistent with
stellar radiation.  This suggests some additional heating mechanism
was operating prior to, and during, 2004.  The most likely mechanism
would appear to be associated with collisional heating within the wind
as dust and gas velocities equalise and turbulence dissipates.  The
kinetic energy in the outflow is more than sufficient to heat the
material to the observed temperature.  The flux density at $5 \mu$m
dropped by a factor of $\sim 18$ in the 14~months between observations
in 2004 and 2005.  This decline corresponds to a drop in blackbody
temperature of $\sim 80$~K, from 320~K in 2004 to $\sim 240$~K in
2005, comparable with $\ltsimeq 200$~K determined by \citet{Eva06}
from $Spitzer$ observations at 20~$\mu$m in 2005 April.  This is a
much larger drop than in the years just prior to and subsequent to
2004-2005 [Figure~\ref{fig-co1-5}\@(a)] and may indicate the rapid
weakening of this additional heating process.

\section{Conclusions}
\label{sec-conclusions}

Observations in the $1-5 \mu$m region show weakening of the CO and
continued cooling of the dust surrounding Sakurai's Object between
2003 and 2007, with a particularly marked temperature decrease between
2004 and 2005.  The 2004 dust temperature is $\sim 200$~K hotter than
can be accounted for by stellar radiation alone.

From an echelle spectrum of CO absorption features around $4.7 \mu$m,
we determine a wind velocity of $500 \pm 80$~km~s$^{-1}$, which we use
to estimate an outer gas shell radius of $9.9 \times 10^{10} \leq
R_{\rm CO} \leq 1.1 \times 10^{11}$~km ($1.4 \times 10^{5} \leq R_{\rm
  CO} \leq 1.6 \times 10^{5}$~R$_\odot$) in 2004.  With consistent gas
and dust temperatures ($320 \pm 30$ and $320 \pm 40$~K, respectively),
we believe the two components to be at least partially mixed.
Assuming continuous, uninterrupted expansion of the CO away from the
central star at constant velocity, the gas shell radius would have
reached $1.5 \times 10^{11} \leq R_{\rm CO} \leq 1.6 \times
10^{11}$~km ($2.2 \times 10^{5} \leq R_{\rm CO} \leq 2.3 \times
10^{5}$~R$_\odot$) by the time of the most recent observation of
Sakurai's Object in 2007 June.  We measure a velocity of peak
absorption of $-270 \pm 15$~km~s$^{-1}$, comparable with $-290 \pm
30$~km~s$^{-1}$ obtained by \citet{Eyr04} for 2003.

By modelling the CO fundamental in a spherically symmetric shell of
material surrounding Sakurai's Object, we have determined an isotopic
ratio of $^{12} \rm C/^{13} \rm C = 3.5^{+2.0}_{-1.5}$.  Application
of a simple isothermal slab model of absorbing CO also yields $^{12}
\rm C/^{13} \rm C \approx 3$.  These results are consistent with
earlier estimates of this isotopic ratio in this system
\citep{Asp99,Pav04,Eyr04,Eva06}.  In particular, agreement with the
atmospheric simulations of \citet{Pav04} demonstrate this low ratio as
being consistent with material having been ejected from the central
star during the VLTP event and being swept out with the fast wind,
rather than forming part of the old PN, or existing in the intervening
interstellar medium.

\section*{Acknowledgements}
\noindent HLW acknowledges studentship support from the University of
Central Lancashire.  MTR acknowledges support from the University of
Central Lancashire.  TRG is supported by the Gemini Observatory, which
is operated by the Association of Universities for Research in
Astronomy, Inc., on behalf of the international Gemini partnership of
Argentina, Australia, Brazil, Canada, Chile, the United Kingdom, and
the United States of America.  Some of the data reported here were
obtained as part of the UKIRT Service Programme.  The United Kingdom
Infrared Telescope is operated by the Joint Astronomy Centre on behalf
of the Science and Technology Facilities Council of the UK.

\bsp

\label{lastpage}
\end{document}